\documentclass[epj]{svjour}

\usepackage{amsmath,amssymb}
\usepackage{graphicx}

\begin{document}

\title{Impact of mobility structure on optimization of small-world networks of mobile agents}
\titlerunning{Mobility patterns and networks of mobile agents}

\author{Eun Lee\inst{1} \and Petter Holme\inst{1}}

\institute{Department of Energy Science, Sungkyunkwan University, Suwon 440-746, Korea}
\mail{holme@skku.edu}
\date{Received: date / Revised version: date}

\abstract{
In ad hoc wireless networking, units are connected to each other rather than to a central, fixed, infrastructure. Constructing and maintaining such networks create several trade-off problems between robustness, communication speed, power consumption, etc., that bridges engineering, computer science and the physics of complex systems. In this work, we address the role of mobility patterns of the agents on the optimal tuning of a small-world type network construction method. By this method, the network is updated periodically and held static between the updates. We investigate the optimal updating times for different scenarios of the movement of agents (modeling, for example, the fat-tailed trip distances, and periodicities, of human travel). We find that these mobility patterns affect the power consumption in non-trivial ways and discuss how these effects can best be handled. 
}


\maketitle

\section{Introduction}

We are living in an era with an increasing amount of wirelessly linked devices. These could be mobile phones (or other devices for human communication), environmental sensors, military appliances, etc. There is an exciting possibility of linking these in a decentralized, self-organized way, without a central control or a pre-assigned connection topology and without the need of building a centralized infrastructure (which takes time, money and planning). Such mobile ad hoc networks (often acronymed ``MANETs''), inspired by distributed systems in nature, could potentially be more robust and energy efficient than traditional architectures~\cite{kumar}. Needless to say, in disasters, military conflicts and other rapidly changing conditions, they would be very useful. Despite this promising outlook, ad hoc networks are not yet in widespread use---for one reason because it is hard to engineer a system to be self organized. The challenge to design ad hoc networks is an interdisciplinary problem uniting engineering and computer science with the natural sciences~\cite{thiemann,hayashiono,krause}. In this work, we will use a physics approach to study a problem in the interface of ad hoc networks and human mobility.

An ad hoc network needs to simultaneously be robust to failures, efficient in transferring information and consume as little energy as possible. The energy consumption is controlled by letting only node pairs conncted by a link relay packets between them. The network topology should also be fair to the agents, so not some agents need to consume much more energy than others. To reduce the energy consumption, the network does not have to relay packets from all other devices within range, a reduced number of them could be enough~\cite{kumar,santi}. The power consumption $P$ of a radio link increases rather fast with the distance $r$ between the devices. In the literature, this relationship has been modeled like $r^\delta$ with $\delta$ ranging between 2 and 4~\cite{liuli}. Note that there is also a cost associated with receiving a packet (that is independent of whether, or not, there is a link to the sender), that we, and most of the literature~\cite{santi}, ignores. The motivation is that this cost is lower and more uniformly distributed than the sending cost.

Given its energy budget, an agent in an ad hoc network would then have a choice to have many short-range connections, or a few long-range connections. More short-range connections would make the network denser and more robust against failures, but it would also decrease the communication efficiency and reliability of the network. The reason is that an information packet then needs to travel more steps to reach from a source to a target. The needed energy to route the packet and the chance of passing a failing device will increase with increasing link density. The good news is that trade-offs need not cost that much. This insight comes from the highly cited work by Watts and Strogatz on the small-world network phenomenon~\cite{wattsstrogatz}. In their model, they start from a locally ordered network and show that you only need a small fraction of long-range connections to obtain a large decrease in the average distance (number of steps from random pairs of source and target). Inspired by this, Holme, Kim and Fodor (HKF) proposed a method to construct the connection topology of an ad hoc network~\cite{hkf}. In their method, each agent has a window of power consumption $[P_\mathrm{min},P_\mathrm{max}]$ and they commit to the network to spend power in that range. Then they fill up their short-range connections except occasionally (controlled by a parameter) when they connect to a more distant agent. (HKF also consider a version of this model where the agents are separated into those only making short- and those only making long-range connections, with similar properties. In this work, we only study the former version.) However, HKF assume that agents are static in space. Even though there are such applications (e.g.\ environmental sensors), this assumption limits the application rather severely.

The HKF protocol can quite straightforwardly be extended to handle mobile agents. One could simply leave the topology intact for some period of time and then reconfigure it according to the protocol. However, in that case, one faces another trade-off problem. Each time one reorganizes the network one needs to spend energy, both for numerical processing and communication. So reorganizing too often is not preferable. Reorganizing too rarely is also not an option since the movement of the agents will deteriorate the original optimization---the energy consumption will increase and there will be a heightened chance of links breaking because devices go out of range from each other. In this paper, we investigate this trade-off problem---how often should one reconfigure the network and how does this depend on the structure of the movement patterns. There are many models for human mobility patterns, ranging from parsimonious to elaborate~\cite{camp}. We take the former approach and investigate three stylized models capturing some simple empirically verified mobility patterns. One uniformly random scenario, mostly as a reference model; one L\'evy flight scenario (where agents take random steps of power-law distributed length); and one pattern mimicking periodic commuters. Both the last two scenarios are claimed to be typical in human mobility~\cite{brockmann,parkleegonzalez,gonz,freeman}.

In the remainder of this paper, we will go through the precise definition of the protocol and the mobility models; present the numerical results and, finally, put these in a wider perspective.

\begin{figure*}
\begin{center}
\includegraphics[width=0.67\linewidth]{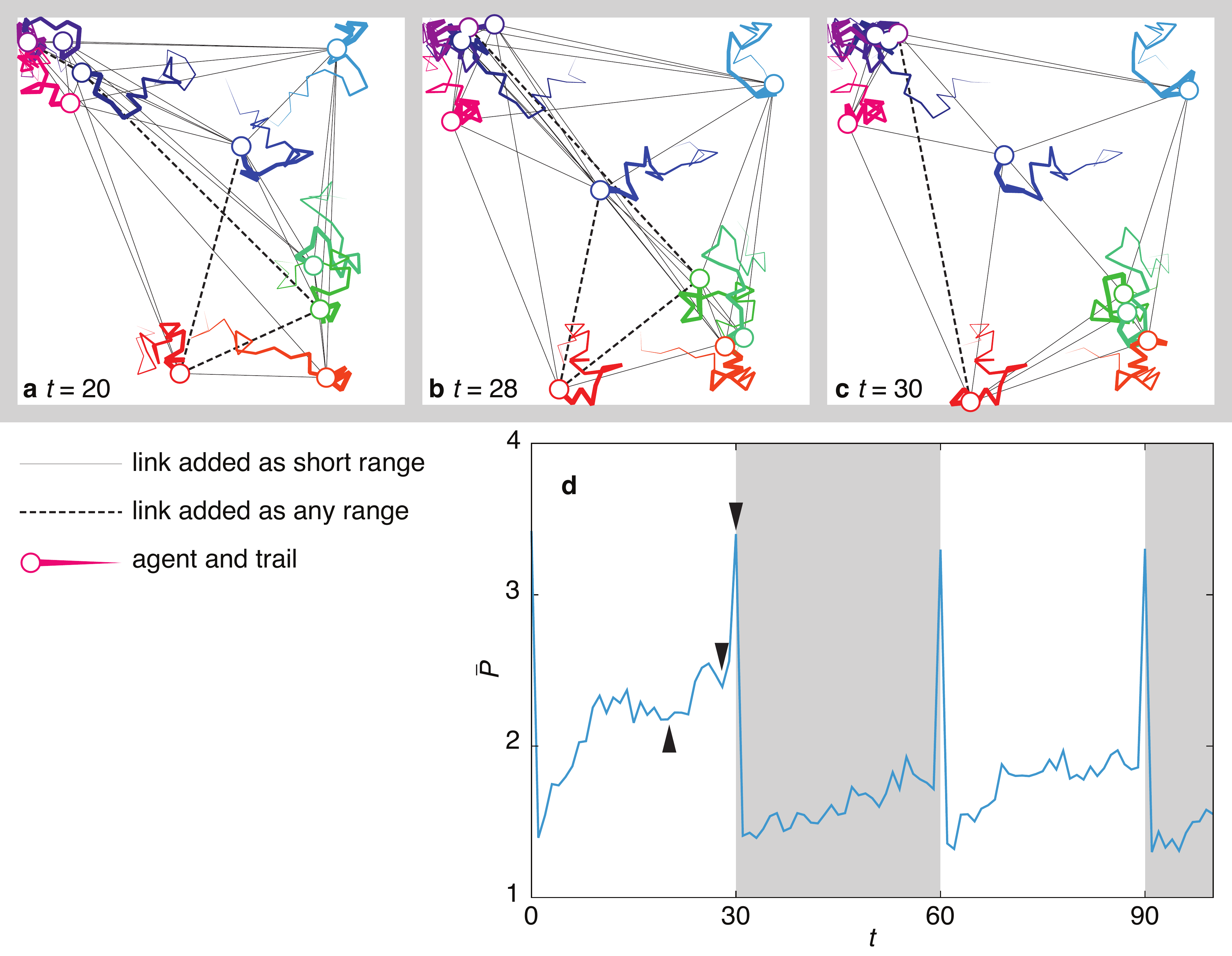}
\end{center}
\caption{An example run of our random walk model of human mobility. In this example we have $N = 10$ agents and updating time $D_U = 30$ and the ratio of any-range connections $q = 0.1$. In the upper panels, we show snapshot configurations at $t = 20$ (a), 28 (b) and 30 (c). These are from the run whose power consumption is displayed at the bottom (the times of the snapshots are indicated by triangles). Dashed lines represent range connections that are unrestricted in length (at the time of the network construction) and the black line show connections added to nearby nodes. The recent trajectories of the nodes are indicated by tails (narrowing with time).}
\label{f1}
\end{figure*}

\begin{figure*}
\begin{center}
\includegraphics[width=0.67\linewidth]{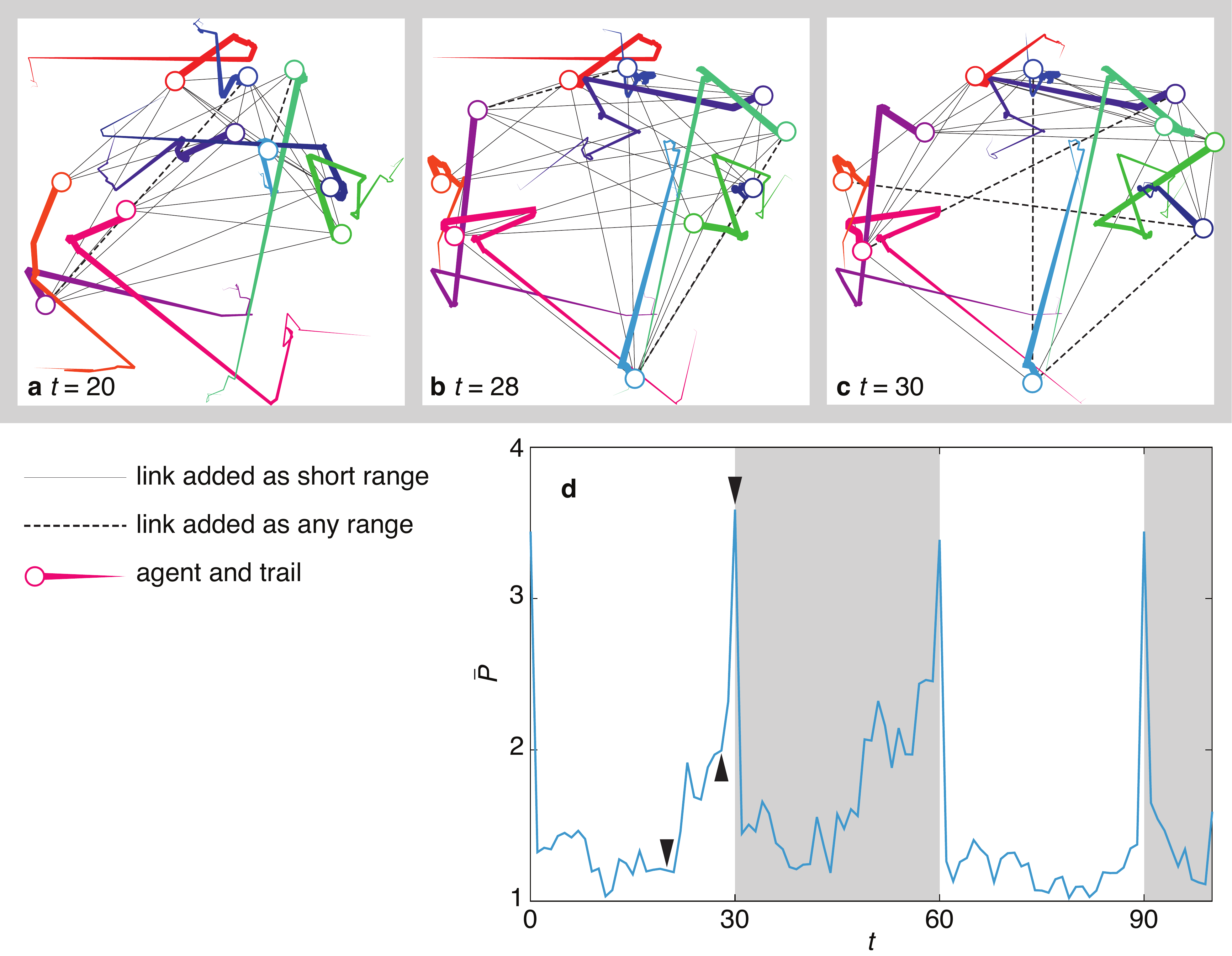}
\end{center}
\caption{An example run of the L\'evy flight model. The simulation setup and details of the plots are the same as in Fig.~\ref{f1}.}
\label{f2}
\end{figure*}

\begin{figure*}
\begin{center}
\includegraphics[width=0.67\linewidth]{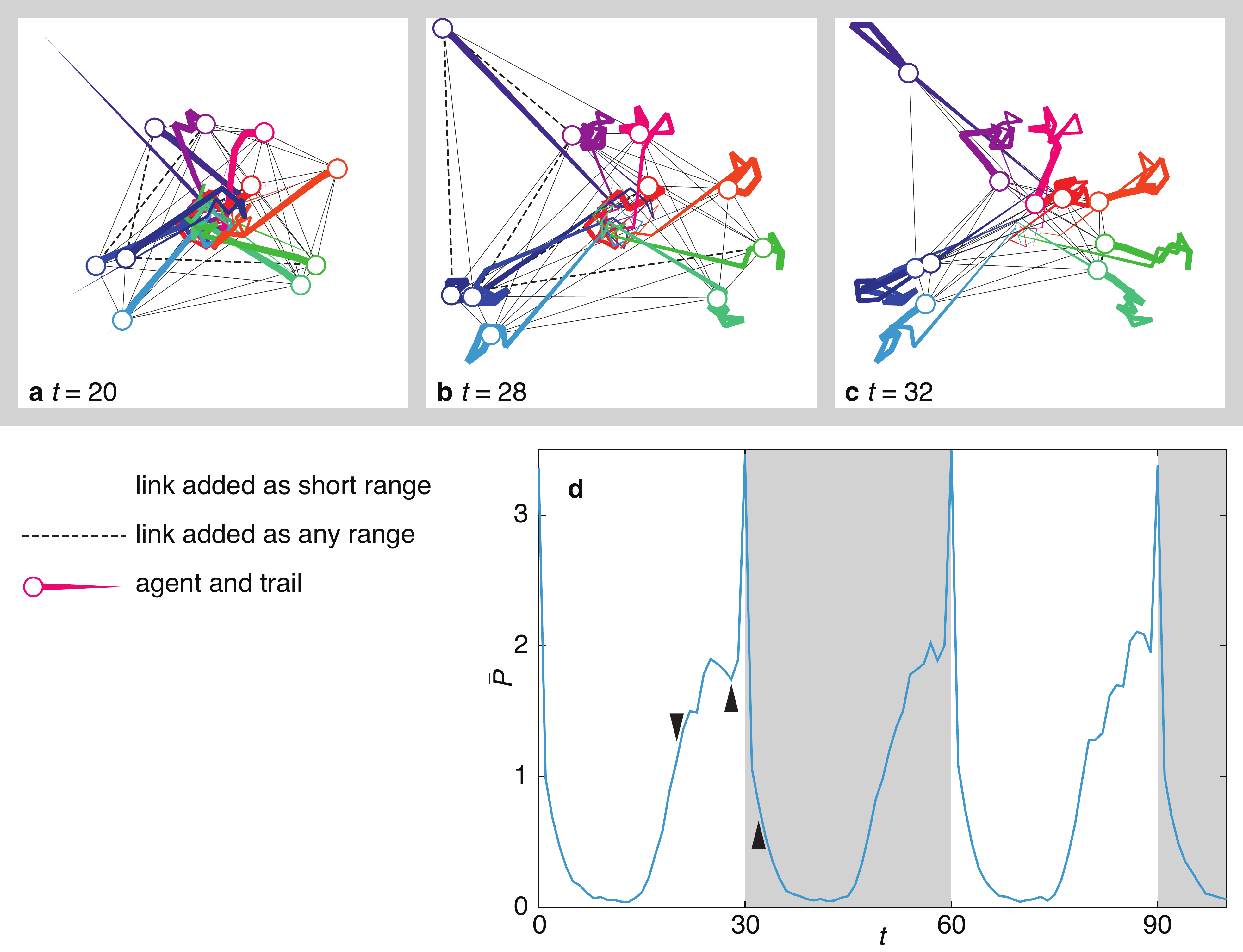}
\end{center}
\caption{An example run of the commuter model. The simulation setup and details of the plots are the same as in Figs.~\ref{f1} and \ref{f2}. The periodicity of the commuting behavior $2D_C$ is the same as the periodicity of the updating $D_U$. The commuting behavior is also synchronized so that agents start moving inwards at the time of the updating.}
\label{f3}
\end{figure*}

\section{Simulation preliminaries}

\subsection{Overview of the simulation}

Our problem is, as described in the Introduction, by nature very complex. Still we try to keep the simulation as simple as possible, at the expense of some realism. That much said, we do believe some of our insights extends to real routing protocols. Our first simplifying assumption is that the network is maintained no matter where the agents are located---if they are far away, the connection just costs more power. As we do not want to relax too many assumptions of Ref.~\cite{hkf}, we let our model inherit this aspect. In a further study, it would be natural to impose a restricted connection range. We also assume agents never join or leave the system. Probably the conclusions would be qualitatively the same of we had a turnover of agents (usually called ``churn'' in the engineering literature). We also assume the agents are confined to the unit square. It is reasonable to assume a practical ad hoc network is confined to a limited region of space~\cite{kumar}, then the particular shape of the region should not matter much as long as the connection range of the devices is longer than the longest distance within it.

Furthermore, we assume the network of radio links are updated at regular intervals (with a time $D_U$ between each update). This set up is similar to real implementations of ad hoc networks~\cite{kumar}. At each update, the connection network is re-optimized according to the HKF protocol~\cite{hkf} (discussed in detail below) and then kept constant until the next update. Throughout the simulations the agents are moving according to a mobility model. We discuss these models below.

\subsection{The Holme-Kim-Fodor model}

Given a static point pattern of agents, and the further assumptions mentioned above, we construct a network of radio connection by the Holme-Kim-Fodor model~\cite{hkf}. Assume an agent $i$ has position $(x_i,y_i)$ and is connected to $k_i$ other agents $\Gamma_i$. Then the power consumption of $i$ is
\begin{equation}\label{eq:pi}
P_i=\sum_{j\in\Gamma_i}\left[(x_i-x_j)^2+(y_i-y_j)^2\right]^{\delta/2} ~,
\end{equation}
where the exponent $\delta$ captures the distance dependence of the radio device. $\delta$-values in the range $[2,4]$ has been used in the literature. Following Ref.~\cite{hkf}, we use $\delta = 2$. This equation neglects the cost of processing incoming packets. This is standard in the literature~\cite{santi} and can be motivated by it contributing less to the energy cost than sending packets, and also more equal for different agents (especially in a setting with mobile agents).

For fairness, we assume an interval $[P_\mathrm{min},P_\mathrm{max}]$ of power consumption such that all agents are guaranteed not to have to spend more battery power than $P_\mathrm{max}$ and they are required to contribute to the network connectivity with at least the power $P_\mathrm{min}$.

The idea of Ref.~\cite{hkf}, is to find a trade-off between battery consumption and quick and efficient communication (short network distances) by mixing cheap short-range connections and more expensive long-range connections. The inspiration comes from the small-world network model of Ref.~\cite{wattsstrogatz} where the authors show that only a small fraction of the links need to be long-range---or more accurately ``any-range'', since they can by construction, also can be short---for the entire network to have logarithmically scaling distances (i.e., being efficient, or ``small world''). In the HKF model, the fraction of any-range links is controlled by the parameter $q$. The construction starts from all agents having no links and proceeds as follows.
\begin{enumerate}
\item Pick a random node $i$ with $P_i < P_\mathrm{min}$.
\item With a probability $q$, pick a random node $j$. If $j$ is distinct from $i$ and not a member of $\Gamma_i$ and if connecting $i$ and $j$ would not make $P_i$ or $P_j$ exceed $P_\mathrm{max}$, then add a link between $i$ and $j$.
\item Otherwise (on average a fraction $1 - q$ of the times), attach $i$ to the closest other agent that is not already a member of $\Gamma_i$. (This could bring the power consumption of $i$ or $j$ above $P_\mathrm{max}$, but for a dense enough system, probably not much, so we allow such $P_\mathrm{max}$ breaches.)
\item Continue until all nodes have a power consumption above $P_\mathrm{min}$.
\end{enumerate}
In our simulations, as the agents move, their power consumption can increase over $P_\mathrm{max}$. We also add a per-agent energy cost $E_U$ for the updating itself. The average power for a node over one updating period of $D_U$ time steps (we take time step, without loss of generality, to be one unit of time) is thus
\begin{equation} \label{eq:barpi}
\bar{P_i} = \frac{1}{D_U}\left[ E_U + \sum_{t=1}^{D_U} P_i(t)\right] ~.
\end{equation}
The total energy consumption of a node throughout an updating period (of duration $D_U$) is the sum of the power consumption for all the time steps according to Eq.~\ref{eq:pi} plus $E_U$. In our simulations we use $P_\mathrm{min} = 1$ and $P_\mathrm{max} = 2$ (just like in Ref.~\cite{hkf}), and $E_U = 2$. We let $\bar{P}$ denote the power consumption averaged over all agents. We run the simulations for $200D_U$ time steps and start taking averages after one updating period (to eliminate effects of the initial condition). We repeat this procedure $n=50$ times for averages. Unless otherwise stated, we use $N = 50$ agents (or nodes in the networks) and $D_U = 5$.

\subsection{Mobility models}

Now we turn to our three mobility models. The first one is a random walk (RW) with uniform step lengths. This is conceptually the simplest possible model and a good null model to compare the other models against. We let each agent take a step of length 0.05 in a random direction. For all three mobility models, we use reflective boundary conditions and initialize the agents to uniformly random coordinates in the unit square.

From this simplest mobility model, we generalize in two different directions. Our first generalization is as a L\'evy flight (LF), where we use a more realistic distribution of the step lengths $\Delta$, namely sampled from a power-law distribution~\cite{chong}.
\begin{equation} \label{eq:powerlaw}
P(\Delta)\propto\Delta^{-1.6}\mbox{~for~} \Delta\in[0.0309\dots,1] ~.
\end{equation}
We take the value of the exponent from Ref.~\cite{brockmann} (even though they study movements of a larger scale than the radio links we have in mind). To facilitate comparison, we let the LF model have the same average step length (0.05), like the other models. We control this by setting the lower cut-off to $0.0309\dots$ (optionally one could have tuned the exponent). The larger cut-off comes from the size of the area---we ignore steps that would take us out of it. Except the step lengths, this model is identical to RW. Second, we mimic the daily routines typical for human life~\cite{jo2012circadian,parkleegonzalez,freeman}. We call it the commuter model (CM). We assume agents move with the same step length (0.05) as RW but alternately (straight) in toward the center $x = y = 0.5$ or (straight) out toward randomly selected ``home'' locations. New home locations are selected at the beginning of the simulation (so our results are averages of $n$ sets of home locations). They move outwards and inwards over intervals of the duration (equaling half the period time) $D_C$. When they are closer than 0.1 from the target (either the center or their home location) they take random steps of length 0.05 within the circle of radius 0.1 from the target. At the start of the simulation (i.e.\ at the first construction of the network), we let the agents be at the home location. This will make the effects of $D_C$ more visible than if one would average the phase difference of the two periodic events, or start at some other relative phase.

\section{Results}

\subsection{Example runs}

In Fig.~\ref{f1}, we show an example run for the RW mobility model. We use $N = 10$, $D_U = 30$, and $q = 0.1$. Three snapshots of the configurations are shown in panels a, b and c. The recent history of the agents is shown by trails decaying in width with age. We can see that, by time $t = 20$ and 28 the network has already deteriorated so that the power consumption has increased to around two-fold of what it was after the first construction. You can also see the original, short-range links are stretched out to the same length as the any-length links. At $t = 30$, the network reconstructed and the power consumption reduced from around 2.4 to around 1.4 (if we disregard the energy cost $E_U$ of the reconstructing the network, which causes the spikes in Fig.~\ref{f1}c). After the reconstruction the trend of the power consumption is increasing, but there are rather large fluctuations due to the small number of agents.

Figure~\ref{f2} shows the corresponding figure to Fig.~\ref{f1} for the LF mobility model. We can see that, even though the fluctuations are larger and more intermittent in the LF model, the general evolution of the system is similar to the RW model. The behavior for the CM model (seen in Fig.~\ref{f3}) is, on the other hand, rather different. After an updating, the power consumption continues to decrease. The reason for this is that, in this case, the commuting behavior is synchronized with the updating so that the system is updated when the agents are close to their homes. After the updating, the agents start moving towards the center, which decreases their distances and reduces the power consumption. Half the time between the network updates, the agents start moving to their home locations. Therefore, during this second half of the interval, the distances and the power consumption increases with time.

\begin{figure}
\begin{center}
\includegraphics[width=0.85\linewidth]{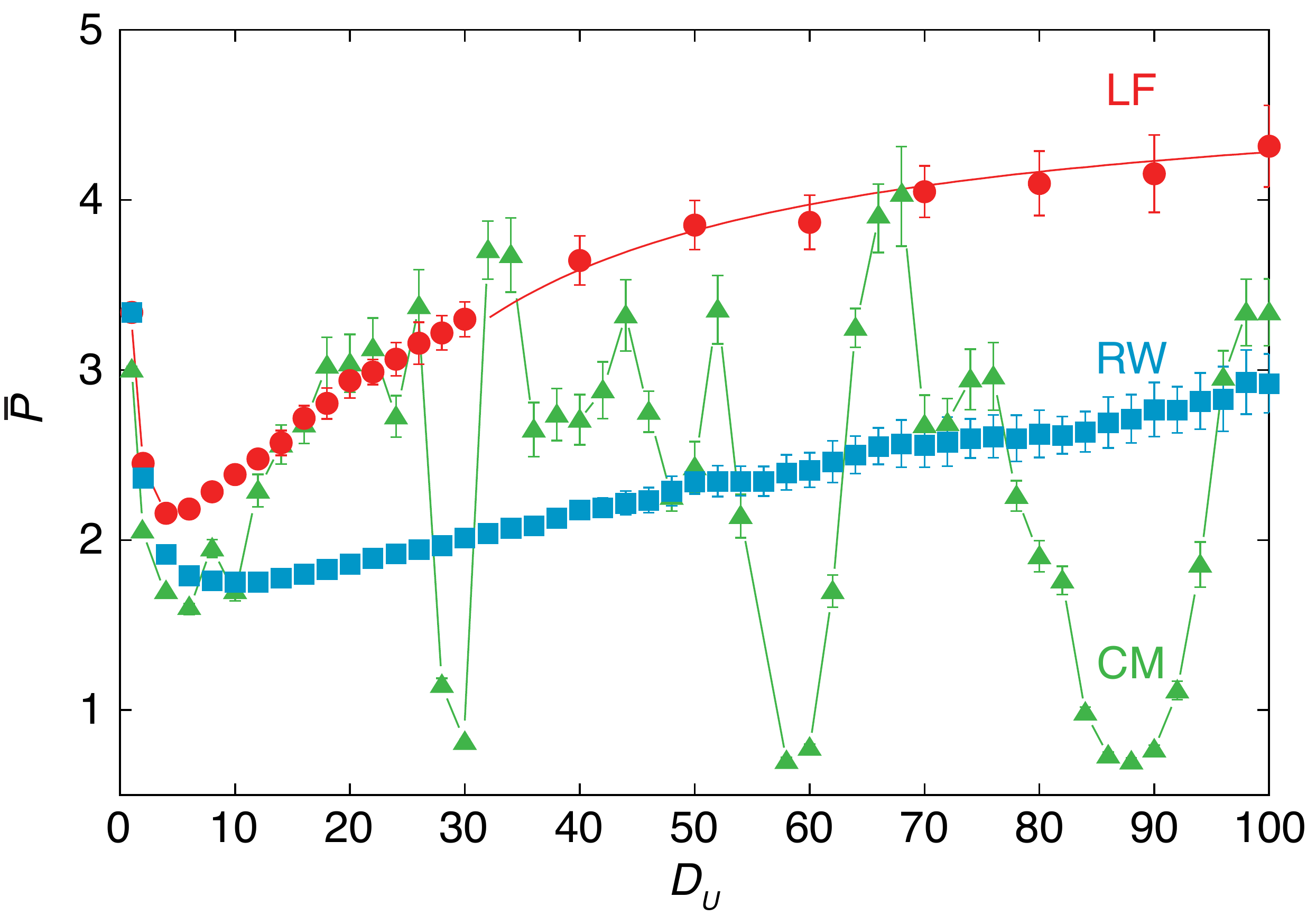}
\end{center}
\caption{Average power consumption for different duration of the updating intervals for the three human mobility models. Error bars represent the standard error. The line is a fit to the form given by Eq.~\ref{eq:piapproxassymp}: $\bar{P}\approx(4.4\pm 0.1) + (9.7\pm 0.2)/D_U$.}
\label{f4}
\end{figure}

\subsection{Effects of the updating time}

Next we turn to the effects of the updating time $D_U$. In Fig.~\ref{f4}, we plot the average power consumption for our three mobility models as functions of $D_U$. For the LF and RW models, there is an early minimum and a slow increase. The minimum represents the time when the energy gain from the optimized configuration balances the cost of the updating. For too short $D_U$, one spends too much energy on constructing the network; for too long $D_U$ the effects of the gain from the optimized configuration gets diluted by the average power-consumption (of what effectively is a random network of the same number of links as the one obtained from the HKF model).

Assume that we can divide an updating interval into a time $\tau_O$ when there is an still an power-lowering effect from the optimization, so the rest of the time there is no other effect of the hkf model optimization than the number of links. Let $E_O$ be the energy consumed during the  power-lowering time. Furthermore, let $P_R$ be the average power consumption when there is no correlation between the links and the position of the agents. Then, for updating times longer than the optimum, the average power consumption is
\begin{equation} \label{eq:piapprox}
\bar{P}\approx \frac{E_U+E_O+P_R(D_U-\tau_O)}{D_U}
\end{equation}
this means $\bar{P}$ will approach $P_R$ like
\begin{equation} \label{eq:piapproxassymp}
\bar{P}\approx P_R - \frac{\mbox{const.}}{D_U}~.
\end{equation}
In practice, we can see this functional for fits reasonable for $D_U > 30$ (Fig.~\ref{f4}). For RW one could only achieve a reasonable fit with our data for $D_U > 80$ (not shown). Once again, this analysis is sketchy (like the assumption $\tau_O$ is a well-defined time scale, the same for all runs).

The CM model shows a radically different behavior. The reason is of course that the power needed to keep the radio links operating varies with the position of the agents---whether they are closer the center or their home locations. Since we (in Fig.~\ref{f4}) construct the network when the agents are at the maximal separation, this leads to the sparsest possible network (the agents prioritize keeping the connectivity at the expense of performance according to the HKF protocol). When the agents move closer to the center the network will consume less power, on average, than the planned lower limit $P_\mathrm{min}$. This pattern repeats for $D_U$ being a multiple of the periodicity of the movement ($D_C = 30$ in Fig.~\ref{f4}). For higher multiplicities ($D_U = 60$ and 90) the valleys become broader. We do not have a simple explanation for neither this phenomenon, nor the minor peaks at other update intervals. If the phase of the commuting movement is chosen at random this structured pattern gets smoothed out to a curve similar to RW and LF.

\subsection{Network structure and optimal updating time}

\begin{figure}
\begin{center}
\includegraphics[width=0.85\linewidth]{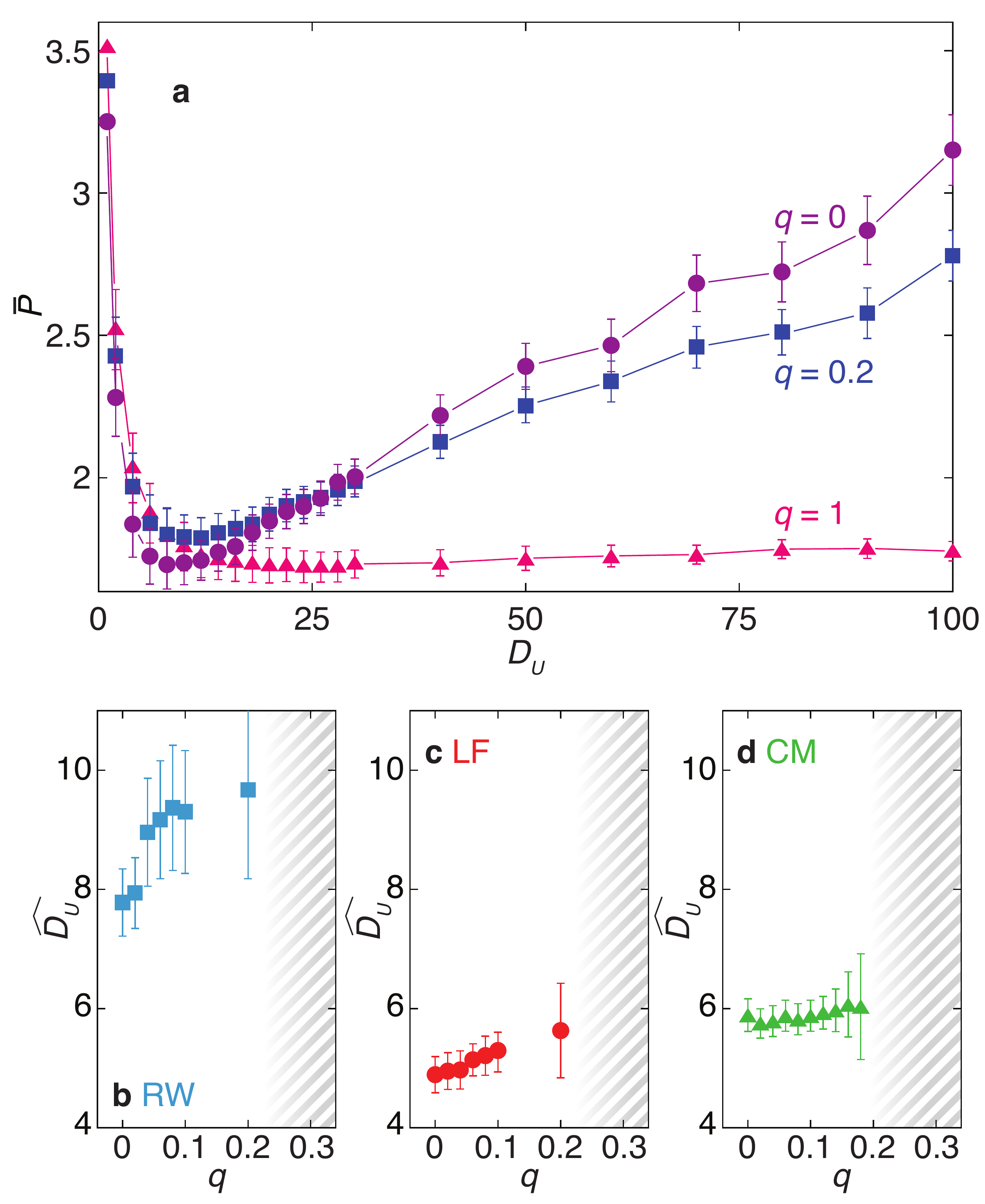}
\end{center}
\caption{The updating duration minimizing the power consumption as a function of the fraction of any-range links. In the shaded areas there is no intermediate optimum---the cost of updating the network outweighs the benefits (so longer $D_U$ is better). The error bars of panel a represent standard errors, while in b--d they represent the standard deviation.}
\label{f5}
\end{figure}

Next we turn to investigating how the optimal updating time depends on the parameters of the HKF protocol. In particular, we investigate the dependence on the fraction of any-range links $q$. See Fig.~\ref{f5}. In panel A, we show curves (of the same type as Fig.~\ref{f4}) for the RW mobility model for some select $q$-values. For low $q$, which is the natural configuration, there is a pronounced minimum (as we have also seen in Fig.~\ref{f4}). For larger $q$-values this minimum becomes wider, and eventually any long enough updating interval would minimize the average power consumption. The reason for the larger $q$ is that the network will be sparser (since the HKF protocol prioritizes connectivity over robustness and efficient communication). In the extreme situation of $q = 1$, the locations of the agents do not matter at all, so updating the network does not matter either. 

Note that this does not mean a large $q$ is necessarily preferable. In the present work, we take the HKF model as the starting point, which is a model optimizing $q$ for robustness and performance given an energy budget. These quantities are optimized for and intermediate $q$-values~\cite{hkf} and are not changed as agents move after the update. In a situation when energy consumption needs to be optimized, and only a minimum robustness and performance is required, one would need to optimize $q$ and $D_U$ simultaneously.

In panels b, c and d of Fig.~\ref{f5}, we plot the optimal updating time $\widehat{D_U}$ as a function of $q$. We show the points as long as we can locate the minimum with a precision higher than five time units. As the minimum gets flatter the imprecision of locating it explodes. To highlight this, we plot the standard deviation, rather than the standard error, in panels b--d. We show the region of ill-defined minima---where the valley is so flat compared to the fluctuations, so that we cannot pinpoint the minimum---as shaded in panels b, c and d. For the RW and LF models, the optimal updating intervals grow with $q$. This effect is stronger for RW than LF (which is in contrast to the $D_U$-dependence of $\bar{P}$, Fig.~\ref{f4}, where LF is most sensitive to parameters than RW). The optimal updating interval of CM model increases very slowly. It is fair to call it independent of $q$. We do not have any simple mechanistic explanation for this, but probably $\widehat{D_U}$ is pinned to the periodicity of the commuter movement.

\begin{figure}
\begin{center}
\includegraphics[width=0.85\linewidth]{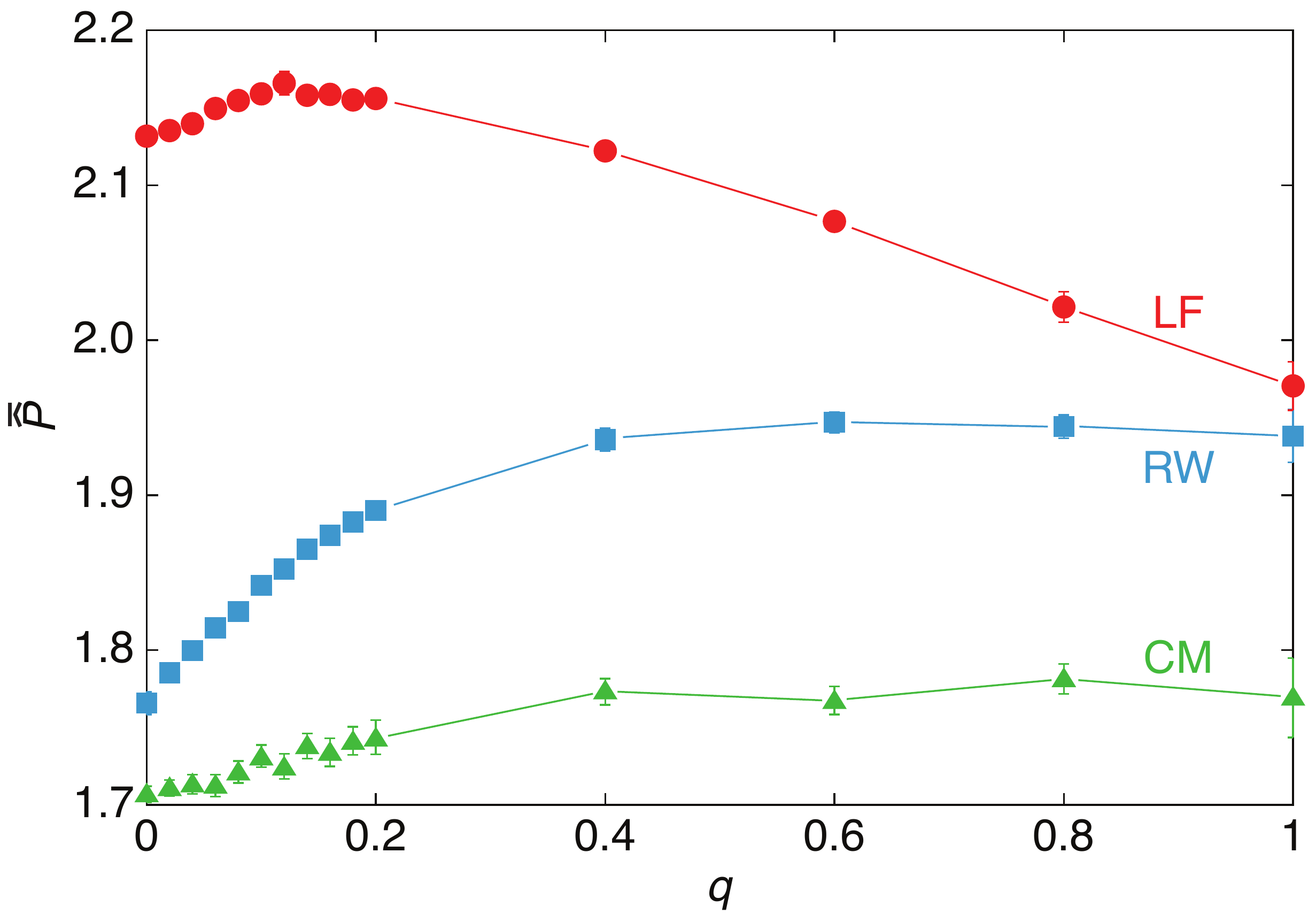}
\end{center}
\caption{Average optimal power consumption as a function of the fraction of long-distance connections $q$ for the three models of human mobility. Error bars represent standard errors.}
\label{f6}
\end{figure}

\subsection{Network structure and optimal power consumption}

Our final analysis concerns the average power consumption at the optimized updating intervals as a function of the fraction of any-range links. The results are shown in Fig.~\ref{f6}. These curves have maxima at intermediate $q$-values. A reason for the increase at small $q$-values is that during the network construction process, one is allowed to use more short-range links which means one can reach over $P_\mathrm{min}$ without overshooting much. Allowing long-range links means that the lastly added link of several nodes is long-range which could bring the power consumption close to $P_\mathrm{max}$ and thus increase the average optimal power consumption. For large $q$-values the effect, as mentioned, is that the network becomes sparser which---as soon as the memory of the original configuration has deteriorated by the movement of the agents---lowers the average power consumption. At $q = 1$, the only difference between the LF and RW is that the larger fluctuations of LF deteriorates the optimization faster (so LF's value becomes a bit higher). The power consumption for the CM model is lower than both RW and LF since, it the links shrink when the agents are close to the center and the DC is a multiple of $\widehat{D_U}$, so that the network construction takes place when the network is as spread out as possible. This effect would, of course, change if we initially constructed the network at some other phase of the agents' movement.

\section{Discussion}

In this paper, we have explored the impact of the movement patterns of agents on the power consumption of wireless ad hoc networks. This is a question about engineering emergent phenomena and as such it is connected to many fields---electric engineering, the theory of distributed computing, and aspects of the physics of complex systems such as small-world networks and L\'evy flights. Roughly speaking our objective is to simultaneously achieve five properties:
\begin{enumerate}
\item Low global power consumption.
\item A fair distribution of the power consumption.
\item A short hopping distance.
\item A network robust to failures and attacks.
\item Parsimony.
\end{enumerate}
In a practical implementation, other issues need to be added to this list, like the computational efficiency and robustness of the algorithms~\cite{kumar,guiliu}. Of the five points above, points 3 and 4 only relate to the static network topology, so we focus on 1 in this work. (While points 2 and 5 are directly taken care of by the network-construction protocol.)

We find that the movement patterns do affect when the network need to be updated in non-trivial ways. The network need to be updated most rarely with the RW mobility model, and most frequently with the LF model (Fig.~\ref{f5}). In other words, heterogeneities of the step lengths induce a need for frequent updating (even though the agents have the same average speed in all simulations). Both RW and LF have increasing optimal updating intervals as functions of the fraction $q$ of any-range links (that tunes the balance between robustness and efficient communication in the network). The CM model, on the other hand, has the same optimal updating interval regardless of $q$. Another finding (Fig.~\ref{f6}) is that the most efficient power consumption of both the RW and LF mobility models have peaks as functions of $q$. Interestingly, the minimum average power consumption happens at opposite extremes for RW and LF ($q = 0$ for RW and $q = 1$ for LF). Also for this analysis, the CM curve is flatter than the others.

There are, of course, many other directions to explore in such a complex problem as the interaction between ad hoc wireless network and mobile agents. Perhaps the least realistic assumption in our work is that any pair of devices can establish a radio link throughout the studied area. In a situation where the agents could move outside of the range of their devices, the overall power consumption (as in Fig.~\ref{f4}) would not be the most critical performance metric. One would also need to evaluate the connectivity of the network, or the average graph distances (cf.\ Ref.~\cite{lenders}). For RW and LF, these functions are monotonically decreasing and the overall picture of our analysis would be the same. For CM there could be a situation where the connectivity and distance metrics oscillates with the periodicity of the movement. We could add another layer of realism by allowing the population to fluctuate. This would, however, probably not alter our conclusions, at least if the connections are not so expensive that the connectivity of the entire system is an issue.

We think there are many interesting future questions relating to the present work. Testing other mobility models would be one promising direction. One could for example assume the agents' movement was affected by their role in the network. If agents tried to (selfishly) minimize their power consumption, or (unselfishly) optimize the communication efficiency and robustness of the network, that could give rise to a yet more complex behavior than we have seen. Another interesting extension of our work would be to use empirical data for human mobility~\cite{vukadinovic,thiemann} instead of models of the mobility, or adapt it for vehicular traffic~\cite{zhuli,thiemann}.

\subsection*{Acknowledgments}

PH was supported by Basic Science Research Program through the National Research Foundation of Korea (NRF) funded by the Ministry of Education (2013R1A1A2011947).


\begin{thebibliography}{10}

\bibitem{brockmann}
D.~Brockmann, L.~Hufnagel, and T.~Geisel.
\newblock The scaling laws of human travel.
\newblock {\em Nature}, 439:462--465, 2006.

\bibitem{camp}
T.~Camp, J.~Boleng, and V.~Davies.
\newblock A survey of mobility models for ad hoc network research.
\newblock {\em Wirel. Commun. Mob. Com.}, 2:483--502, 2002.

\bibitem{freeman}
M.~P. Freeman, N.~W. Watkins, E.~Yoneki, and J.~Crowcroft.
\newblock Rhythm and randomness in human contact.
\newblock In {\em 2014 IEEE/ACM International Conference on Advances in Social
  Networks Analysis and Mining (ASONAM 2014)}, pages 184--191, Los Alamitos,
  CA, USA, 2010. IEEE Computer Society.

\bibitem{gonz}
M.~A. Gonz\'alez, C.~Hidalgo, and A.-L. Barab\'asi.
\newblock Understanding individual human mobility patterns.
\newblock {\em Nature}, 453:779--782, 2008.

\bibitem{guiliu}
J.~Gui and A.~Liu.
\newblock A new distributed topology control algorithm based on optimization of
  delay and energy in wireless networks.
\newblock {\em J. Parallel Distrib. Comput.}, 72(8):1032--1044, Aug. 2012.

\bibitem{hayashiono}
Y.~Hayashi and Y.~Ono.
\newblock Geographical networks stochastically constructed by a self-similar
  tiling according to population.
\newblock {\em Phys. Rev. E}, 82:016108, Jul 2010.

\bibitem{hkf}
P.~Holme, B.~J. Kim, and V.~Fodor.
\newblock Heterogeneous attachment strategies optimize the topology of dynamic
  wireless networks.
\newblock {\em Eur. Phys. J. B}, 73(4):597--604, 2010.

\bibitem{jo2012circadian}
H.-H. Jo, M.~Karsai, J.~Kert{\'e}sz, and K.~Kaski.
\newblock Circadian pattern and burstiness in mobile phone communication.
\newblock {\em New J. Phys.}, 14(1):013055, 2012.

\bibitem{krause}
W.~Krause, J.~Scholz, and M.~Greiner.
\newblock Optimized network structure and routing metric in wireless multihop
  ad hoc communication.
\newblock {\em Physica A}, 361(2):707--723, 2006.

\bibitem{kumar}
S.~Kumar, V.~S. Raghavan, and J.~Deng.
\newblock Medium access control protocols for ad hoc wireless networks: A
  survey.
\newblock {\em Ad Hoc Networks}, 4(3):326--358, 2006.

\bibitem{lenders}
V.~Lenders, J.~Wagner, and M.~May.
\newblock Analyzing the impact of mobility in ad hoc networks.
\newblock In {\em Proceedings of the 2Nd International Workshop on Multi-hop Ad
  Hoc Networks: From Theory to Reality}, REALMAN '06, pages 39--46, New York,
  NY, USA, 2006. ACM.

\bibitem{liuli}
J.~Liu and B.~Li.
\newblock Distributed topology control in wireless sensor networks with
  asymmetric links.
\newblock In {\em Global Telecommunications Conference, 2003. GLOBECOM '03.
  IEEE}, volume~3, pages 1257--1262, Dec 2003.

\bibitem{parkleegonzalez}
J.~Park, D.-S. Lee, and M.~C. Gonz\'alez.
\newblock The eigenmode analysis of human motion.
\newblock {\em Journal of Statistical Mechanics: Theory and Experiment},
  2010(11):P11021, 2010.

\bibitem{chong}
I.~Rhee, M.~Shin, S.~Hong, K.~Lee, and S.~Chong.
\newblock Human mobility patterns and their impact on routing in human-driven
  mobile networks.
\newblock In {\em Proceedings of the Sixth Workshop on Hot Topics in Networks},
  2007.

\bibitem{santi}
P.~Santi.
\newblock Topology control in wireless ad hoc and sensor networks.
\newblock {\em ACM Computing Surveys}, 37(2):164?194, 2005.

\bibitem{thiemann}
C.~Thiemann, M.~Treiber, and A.~Kesting.
\newblock Longitudinal hopping in intervehicle communication: Theory and
  simulations on modeled and empirical trajectory data.
\newblock {\em Phys. Rev. E}, 78:036102, Sep 2008.

\bibitem{vukadinovic}
V.~Vukadinovic, F.~Dreier, and S.~Mangold.
\newblock Impact of human mobility on wireless ad hoc networking in
  entertainment parks.
\newblock {\em Ad Hoc Networks}, 12:17--34, 2014.

\bibitem{wattsstrogatz}
D.~J. Watts and S.~H. Strogatz.
\newblock Collective dynamics of {`small-world'} networks.
\newblock {\em Nature}, 393:440--442, 1998.

\bibitem{zhuli}
H.~Zhu and M.~Li.
\newblock {\em Studies on Urban Vehicular Ad-hoc Networks}.
\newblock SpringerBriefs in Computer Science. Springer, New York, 2013.

\end{thebibliography}
\end{document}